\documentclass[fleqn,twoside]{article}
\usepackage[headings]{espcrc2}

\readRCS
$Id: espcrc2.tex,v 1.2 2004/02/24 11:22:11 spepping Exp $
\usepackage{graphics}
\usepackage{macros}
\usepackage[figuresright]{rotating}

\newcommand{\AmS}{{\protect\the\textfont2
  A\kern-.1667em\lower.5ex\hbox{M}\kern-.125emS}}

\title{Determining fundamental parameters of QCD on the lattice}

\author{Rainer Sommer\address[Desy]{DESY, % \\
        Platanenallee 6, 15738 Zeuthen, Germany}%
        \thanks{This work is strongly supported 
        by the DFG in the
        SFB Transregio 9 ``Computational Particle Physics''. 
        }
	\hfill
	{\onecol{2.8cm}{\vspace{-3cm} \it DESY 06-101 \\ SFB/CPP-06-29}
	\vspace{0.1cm}}
        }
       
\def\cblu{}

\runtitle{Fundamental parameters}
\runauthor{R. Sommer}

\begin{document}

\begin{abstract}
We summarize the status of determinations of fundamental parameters of QCD
by the \includegraphics*[width=10mm]{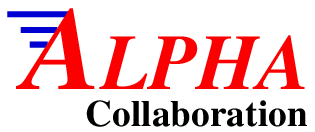}.
\vspace{1pc}
\end{abstract}

\maketitle

\section{INTRODUCTION}
The natural choice for the 
fundamental parameters
of QCD are the renormalization group invariant (RGI) quark
masses, $M_i, i=\mrm{u,d,...}$, and the $\Lambda$-parameter. 
The RGI masses do not depend on a renormalization scheme and the 
scheme dependence of the $\Lambda$-parameter is {\em exactly} given by
a one-loop coefficient. These quantities are in principle free of
perturbative uncertainties.

Their determination from low energy hadron data
is a % noble 
relevant mission for lattice QCD. Before entering details, 
we discuss a few aspects of more conventional 
determinations of $\alpha$ (and $\Lambda$).
One considers experimental observables $O_i$ depending on an
overall energy scale $q$ and possibly some additional kinematical
variables denoted by $y$.
The observables can be computed in a perturbative series
which is usually written in terms of the $\MSbar$ coupling 
$\alphaMSbar=\gbar^2_{\msbar}/4\pi$,
\footnote{
%%%%%%%%%%
We can always arrange the definition of the observables such that
they start with a term $\alpha$. For simplicity we neglect all quark
mass dependencies; they are irrelevant in this discussion.
} \hfill $(\mu=q)$
\bes
 O_i(q,y) = \alphaMSbar(\mu) + A_i(y)\alphaMSbar^2(\mu)+\ldots\,
 \enspace . \label{e_O_i}
\ees
For example $O_i$ may be constructed from jet cross sections and
$y$ may be related to the details of the definition of a jet.

The renormalization group describes the energy dependence
of $\gbar$ in a general scheme,
\bes
  \mu {\partial \bar g \over \partial \mu} &=& \beta(\bar g) \enspace ,
     \label{e_RG}
\ees
here the $\beta$-function has an asymptotic expansion
\bes
 \beta(\bar g) & \buildrel {\bar g}\rightarrow0\over\sim &
 -{\bar g}^3 \left\{ b_0 + {\bar g}^{2}  b_1 + \ldots \right\}
                      \enspace , \nonumber\\
 &&b_0=\frac{1}{(4\pi)^2}\bigl(11-\frac{2}{3}\nf\bigr)
                      \enspace ,\nonumber\\
 &&b_1=\frac{1}{(4\pi)^4}\bigl(102-\frac{38}{3}\nf\bigr) \enspace ,
 \label{e_RGpert}
\ees
with higher order coefficients $b_i, \, i>1 $ that depend on the scheme.

We note that -- neglecting experimental uncertainties -- $\alphaMSbar$
extracted in this way is obtained with a precision given by the terms
that are left out in \eq{e_O_i}. In addition to (say) $\alpha^3$-terms, there
are non-perturbative contributions which may originate
from renormalons, 
instantons or, most importantly, may have an origin that no physicist
has yet uncovered. Empirically, one observes that values
of $\alphaMSbar$ determined
at different energies and evolved to a common reference point
using \eq{e_RG} including $b_2$
agree rather well with each other; the aforementioned uncertainties are
apparently not very large.
Nevertheless, the standard determinations of $\alpha$ are limited in precision
because of these uncertainties and in particular if there was a significant
discrepancy between $\alpha$ determined at different energies and/or processes
it would
be difficult to say whether this was due to the terms left out in \eq{e_O_i}
or was due to terms missing in the Standard Model Lagrangian.

It is an obvious possibility, and at the same time a challenge, for
lattice QCD to achieve a determination of $\alpha$ in one non-perturbatively (NP)
well defined scheme at large $\mu$. There one may use the
perturbative approximation for $\beta(\gbar)$ inserted into 
the exact solution of \eq{e_RG},
\bes \label{e_lambdapar}  \label{e:lambda}
 \Lambda &=&\mu \left(b_0\gbar^2(\mu)\right)^{-b1/(2b_0^2)} \rme^{-1/(2b_0\gbar^2(\mu))} 
        \\  && \times
           \exp \left\{-\int_0^{\gbar(\mu)} \rmd x
          \left[\frac{1}{ \beta(x)}+\frac{1}{b_0x^3}-\frac{b_1}{b_0^2x}
          \right]
          \right\} \enspace , \nonumber
\ees
in order to extract the $\Lambda$-parameter. 

\subsection{Problem}
In an implementation of this idea one needs to satisfy the following criteria.\\
 $\bullet$ Compute $\alpha(\mu)$ at energy scales of
       $\mu\sim 10\,\GeV$ or higher in order to connect in a 
        controlled manner to the perturbative regime.\\
 $\bullet$ Keep $\mu$ removed from the lattice cutoff $a^{-1}$ to
       avoid large discretization effects and to be able to
       extrapolate to the continuum limit.\\
 $\bullet$ 
       Avoid finite
       size effects in the MC simulations by keeping the box size $L$ large
       compared to the inverse of the mass of the pion, $\mpi$.\\

These conditions are summarized by
\bes
  L \, \gg \,{1\over \mpi} \,\sim\,{ 1 \over 0.14\GeV} \,\, \gg \,\,
  {1 \over \mu} \, \sim \, {1 \over 10\GeV}  \,\,  \gg a \enspace ,
  \nonumber % \label{e_conditions}
\ees
which means that one must perform a MC-computation on an $N^4$ lattice
with $N \equiv L/a \gg 70$: an impossible task 
for some time to come.

\subsection{Solution}

\begin{figure}[ht]
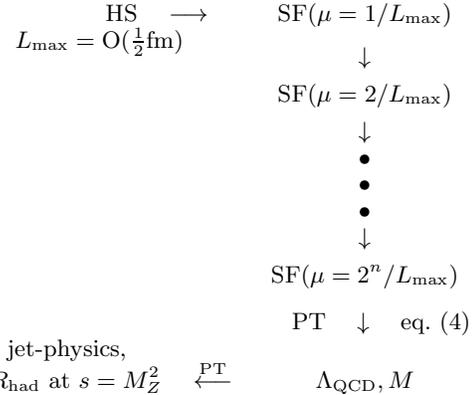

{\small
\bes
 {\rm HS} \quad \longrightarrow \quad
      &{\rm SF} (\mu=1/{ L_{\rm max}})& \quad
               \nonumber \\[-1ex]
{ L_{\rm max}}=\rmO(\frac{1}{2}\fm) \qquad && \nonumber \\[-2ex]
      &\downarrow&  \nonumber \\
      &{\rm SF} (\mu=2/{ L_{\rm max}})&  \nonumber \\
      &\downarrow&  \nonumber \\[-1ex]
      &\bullet&  \nonumber \\[-1ex]
      &\bullet&  \nonumber \\[-1ex]
      &\bullet&  \nonumber \\[-1ex]
      &\downarrow&  \nonumber \\
      &{\rm SF} (\mu=2^n/{ L_{\rm max}})& \nonumber \\[1ex]
   &\quad \mbox{ \small PT} \quad  \downarrow \quad \mbox{\small \eq{e:lambda}} &  \nonumber \\[-1ex]
\hbox{DIS, jet-physics, } \qquad\qquad && \nonumber \\[-1ex]
R_\mrm{had}\mbox{ at } s=M_Z^2 \quad \stackrel{\rm  PT}{\longleftarrow}
        &\Lambda_{\rm QCD}, M & \nonumber
\ees
}
\vspace{-15mm}
\caption{The strategy for a non-perturbative computation of
         short distance parameters. % ``SF'' means \SF scheme,
        % which is the finite volume scheme adapted in the 
        % practical numerical computations.
\label{f_strategy}}
\end{figure}

The simple solution is to identify the renormalization scale and the 
box size $L$ \cite{alpha:sigma}, and step up the energy scale 
recursively. Each step then requires only $L/a\gg1$, see \fig{f_strategy}.
%%%%%%%%%%%%%%%%%%%%%%%%%%%%%FIGURE%%%%%%%%%%%%%%%%%%%%%%%%%%%%%%%%%%%
%%%%%%%%%%%%%%%%%%%%%%%%%%%%%%%%%%%%%%%%%%%%%%%%%%%%%%%%%%%%%%%%%%%%

\section{RUNNING OF THE COUPLING}

We choose a NP defined
\SF renormalization scheme (SF) and set all renormalization
conditions at zero quark mass (see \cite{alpha:nf2}
and references therein). An essential property of this
scheme is the use of Dirichlet boundary conditions in time,
which allow for MC simulations at vanishing quark mass. 

\subsection{The step scaling function}

One vertical step in the figure is achieved as follows: 
we start from a given value of the coupling, $u=\gbar^2(L)$.
When we change the length scale by a factor $2$, the coupling
has a value $\gbar^2(2L)=u'$ .
The step scaling function, $\sigma$,
is then defined as
\bes
 \sigma(u)&=&u' \enspace .
\ees
It is a
discrete $\beta$-function whose knowledge allows for the
recursive construction
of the running coupling at discrete values of the
length scale,
$
  u_k = \gbar^2(2^{-k} \Lmax) \enspace , % \label{e_uk}
$
once a starting value $u_\mrm{max} = \gbar^2(\Lmax)$ is specified. 
The step scaling function has a perturbative expansion
$
 \sigma(u)= u + 2 b_0 \ln(2) u^2 +\ldots  \enspace .
$

On a lattice with finite spacing, $a$, it
has an additional dependence on
the resolution $a/L$. Its computation then
consists of the following steps:\\
 1. Choose an $(L/a)^4$ lattice ($1\leq x_\mu/a=\leq L/a$).\\
 2. Tune the bare coupling $g_0$ such that the renormalized
           coupling $\gbar^2(L)$ has the value $u$ and tune the bare quark mass $\mbare$
           such that the PCAC-mass 
           vanishes.\\
 3. At the same values of $g_0,\mbare$, simulate a $(2L/a)^4$ lattice; compute
           $u'=\gbar^2(2L)$. The lattice step scaling function is
           $\Sigma(u,a/L)=u'$. \\
 4. Repeat steps 1.--3. with different resolutions $L/a$ and extrapolate
           $\sigma(u)=\lim_{a/L \to 0}\Sigma(u,a/L)$. \\[1ex]
Note that step 2. takes care of the renormalization and 3. determines
the evolution of the {\it renormalized} coupling.
The continuum extrapolation 4. is an essential part. In QCD, starting
with the pure gauge theory, a lot of effort -- both perturbative, analytical
and by MC -- has been invested in 
understanding this step properly. We do not have the space here to 
explain it in detail, but as a result of 
\cite{pert:2loop_fin,mbar:pap1,alpha:letter,alpha:nf2,alpha:su2impr} 
and other work cited in these references, 
we are convinced that no systematic lattice errors are left. 

\section{RESULTS}

\begin{figure}[!b]
\includegraphics*[width=50mm]{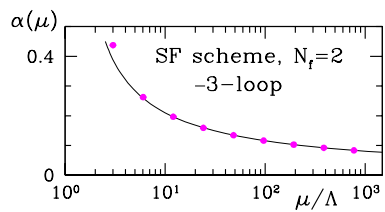}
\vspace{-10mm}
\caption{Running coupling for $\nf=2$.}
\label{f:runn2}
\end{figure}

We proceed directly to discussing the results. Numerical 
data for the step scaling function have been obtained
for 7 values of $u$ in the range of $u=1.0$ to
$u=3.3$ for QCD with $\nf=2$ dynamical flavours~\cite{alpha:letter,alpha:nf2},
while for $\nf=0$ there are 11 points in $u$ between $u=0.9$ and
$u=2.4$ \cite{alpha:su3,mbar:pap1}. After taking the continuum limit,
the extracted $\sigma(u)$ have a precision of
around 0.5\% at the smallest coupling (where it runs slowly) 
to 2\% at the largest one. 
The numerical values of $\sigma(u)$ 
are then represented by 
a smooth interpolating function (a polynomial in $u$). 
With this function the running coupling can be constructed
recursively; the result is shown in \fig{f:runn2} and \fig{f:runn0}.

With a
start value for the $\beta$-function taken from perturbation 
theory (3-loop) at the weakest coupling ($\alpha \approx 0.08$), one can also 
set up a recursion for the $\beta$-function itself \cite{alpha:nf2}.
%%%%%%%%%%%%%%%%%%%%%%%%%%%%%%%%%%%%%%%%%%%%%%%%%
\begin{figure}[t]
\vspace{9pt}
\includegraphics*[width=70mm]{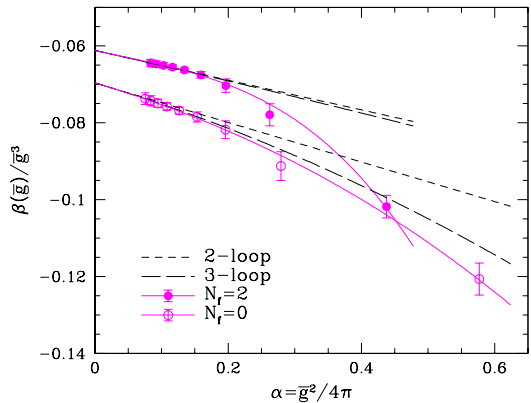}
\vspace{-10mm}
\caption{The QCD $\beta$--function in the SF scheme.}
\label{f:beta}
\end{figure}
%%%%%%%%%%%%%%%%%%%%%%%%%%%%%%%%%%%%%%%%%%%%%%%%%
As \fig{f:beta} shows, its agreement with perturbation theory 
is excellent at weak couplings $\alpha<0.2$, while at
the largest couplings significant deviations
from perturbation theory are present for $\nf=2$. 
Indeed the difference between non-perturbative
points and 3-loop can't be described by an
effective 4-loop term {\em with a reasonable coefficient}.
At the same time the perturbative series just by itself 
does not show signs of its failure at, say, $\alpha\approx0.4$.
Similar deviations from perturbation theory, at even
somewhat smaller values of $\alpha$ have
been found in the NP scale dependence 
of composite operators \cite{4ferm:nf0,zastat:pap3}.

\begin{table}[b!]
{%\footnotesize
\caption{Lambda parameter in units of $r_0$ for $\nf=2$. \label{t:lamr0}} % for
        %different resolutions in the low energy part 
        %of the calculation.}
\begin{center}
\begin{tabular}{ccc}
\hline \\[-1.9ex]
                  & \multicolumn{2}{c}{$u_{\rm max}=4.61$} \\[0.2ex]
  $r_0/a$ &  $L_{\rm max}/a$ & $\Lambda_{\MSbar}\, r_0$ \\[0.5ex]
\hline\\[-1.5ex]
 5.45(5)(20) &  6.00(8)  & { 0.610(25)} \\           
 6.01(4)(22) &  6.57(6)  & { 0.614(24)} \\ 
 7.01(5)(15) &  7.73(10) & { 0.609(16)} \\[0.5ex]     
\hline
\vspace{-8mm}
\end{tabular}
\end{center}
}
\end{table}

We return to the running couplings shown in 
 \fig{f:runn2} and \fig{f:runn0}. In the zero flavour case,
also the region of $\mu$ of around $250\,\MeV$ was 
investigated with a specifically adapted strategy~\cite{lat01:jochen}. 
In this region, the SF coupling 
shows the rapid growth expected from a strong coupling 
expansion.  

Initially, the graphs  \fig{f:runn2} and \fig{f:runn0}
are obtained for $\mu$ in units of $\mu_\mrm{min}=1/\Lmax$
defined by $\gbar^2(\Lmax)=u_\mrm{max}$. One chooses 
 $u_\mrm{max}$ relatively large, but within the range covered
by the non-perturbative computation of $\sigma(u)$.
Since it has been verified that perturbative running sets
in (in the SF scheme) at the larger $\mu$ that were simulated, 
one can use \eq{e:lambda}
with the {\em perturbative} $\beta$-function in
{\em that} region.  In this way $\Lmax$ can be expressed
in units of $\Lambda^{-1}$:
\bes
  \nf=0&&u_\mrm{max}=3.48\nonumber\\[-1ex] &&\ln(\Lambda_{\msbar} \Lmax)= -0.84(8) 
  \label{e:lll0}\\
  \nf=2&&u_\mrm{max}=4.61\nonumber\\[-1ex] &&\ln(\Lambda_{\msbar} \Lmax)= -0.40(7)\,.  
  \label{e:lll2}
\ees
It remains to relate the artificially defined
length scale $\Lmax$ to an experimentally measurable
low energy scale of QCD such as the proton mass or the
kaon decay constant, $\Fk$. So far we have evaluated $\Lmax$ 
in units of the 
scale parameter $r_0$, which has a precise definition
in terms of the QCD static
quark potential \cite{pot:r0} 
and is well calculable on the lattice.
But it is related to experiments
only through potential {\em models}: $r_0\approx0.5\,\fm$. 
For $\nf=0$ the continuum limit is
$\Lmax/r_0=0.738(16)$,    
$\Lambda_{\MSbar}^{(0)} r_0 = 0.60(5)$ \cite{pot:intermed}.

%%%%%%%%%%%%%%%%%%%%%%%%%%%%%%%%%%%%%%%%%%%%%%%%%%
%%%%%%%%%%%%%%%%%%%%%%%%%%%%%%%%%%%%%%%%%%%%%%%%%
In the $\nf=2$ theory, the situation is illustrated in
\tab{t:lamr0} which relies on results for $r_0/a$
from \cite{JLQCD:nf2b52,QCDSF:nf2mstrange}. On the one hand,
all the numbers are consistent, indicating
that lattice spacing effects are small, on the other
hand the first colum shows that $a$ is not yet varied 
very much. At the moment we quote 
$\Lambda_{\MSbar}^{(2)} r_0 = 0.62(4)(4)$, where the
second error generously
covers the range of \tab{t:lamr0} as well as numbers 
obtained with $u_\mrm{max}=3.65$ \cite{alpha:nf2} and the first one 
comes from \eq{e:lll2}.

\section{DISCUSSION}

\begin{figure}[!b]
\includegraphics*[width=50mm]{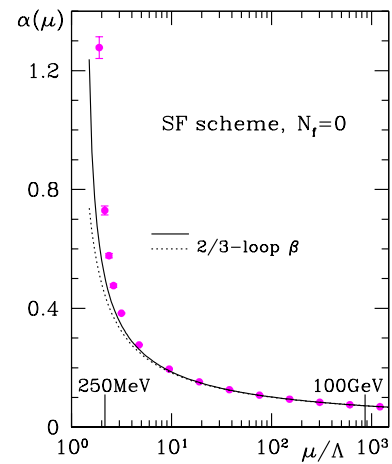}
\vspace{-10mm}
\caption{Running coupling for $\nf=0$.}
\label{f:runn0}
\end{figure}

The scale dependence of the SF coupling is
close to perturbative below $\alpha_\mrm{SF} =0.2$ and becomes
non-perturbative above $\alpha_\mrm{SF} =0.25$. In fact 
a strong coupling expansion suggests that {\em this} 
coupling grows exponentially for large $L$ and 
in the $\nf=0$ theory it was possible to verify this
behavior for $L$ close to $1 \,\fm$ \cite{lat01:jochen} (\fig{f:runn0}).

In the following table, we compare our
results for  $\cblu \Lambda_{\MSbar} r_0$ to selected
phenomenological ones, where we set $r_0=0.5\,\fm$.\\[1ex]
{\footnotesize
  \begin{tabular}{l  l c c c c }
        $\nf$:  &   0  &  2  &  4  &  5 \\ \hline       
        \cite{alpha:nf2,alpha:su3}  &\cblu 0.60(5) &\cblu 0.62(6) & & \\[0.5ex]
        \cite{Bethke:2004uy} & & & 0.74(10) &0.54(8) \\[0.5ex]
        \cite{alpha:blum1}  &&& 0.57(8) & \\
   \hline
  \end{tabular}
}
\vspace{3mm}

There appears to be an irregular $\nf$-dependence, but we note that\\
1.~~the errors are not very small,\\
2.~~the 4-flavour  $\Lambda$ is related to the 5-flavour one
by perturbation theory \cite{alpha:bernwetz} which is
then required to be accurate for  $\mu \ll m_\mrm{beauty}$.

%%%%%%%%%%%%%%%%%%%%%%%%%%%%%%%%%%%%%%%%%%%%%%%%%%
%%%%%%%%%%%%%%%%%%%%%%%%%%%%%%%%%%%%%%%%%%%%%%%%%

\subsection{Improvements of lattice results}
The replacement of $\Lmax/\r_0$ by
$\Fk\times\Lmax$ (computed at small enough 
light quark masses and small $a$)
is the most urgent step and an effort is presently being made.
Also the strange quark sea has to be included (``2+1'') and one should 
estimate the effects of the charm quark. Such 2+1 simulations are
for example being carried out by JLQCD and CP-PACS, who have also 
studied the computation of the SF coupling with the gauge action they 
are using\cite{alpha:Iwasaki1}.

I am convinced that  lattice results will yield the best
controlled and most precise $\Lambda$ in the long run.
The reason is simple: 
I essentially described {\em all} the sources of systematic errors and
the kind of assumptions one has to make are minimal.

There is a large number of other results
from lattice gauge theory, 
where $\alpha(\mu)$  is extracted from a quantity related
to the cutoff: $\mu \sim a^{-1}$. Some of them cite a very small error
\cite{alpha:lepagemilk}. As discussed for example in \cite{reviews:schlad},
it is, however,  not easy to estimate the systematic errors 
in these computations,
mainly because one cannot separate higher order 
perturbative corrections from discretization errors.
It thus remains very desirable to carry out the program of 
\fig{f_strategy} with good precision and the relevant number of flavors.

\section{QUARK MASSES}

% Their computation poses the same basic problem and
The strategy discussed above is easily generalized 
to include the quark masses \cite{mbar:pap1}. A useful ingredient is again
to choose a mass independent renormalization scheme 
(even for $\mbeauty$) where ($i=$ up, down, strange ...)
\bes
  \mbar_i(\mu) = Z_m(\mu a,g_0)\,
                   m^\mrm{bare}_i(g_0)
\ees
with a flavour independent renormalization factor $Z_m(\mu a,g_0)$.
The scale dependence of the running masses can then
be computed  once and for all. The RGI masses % (scale and scheme independent)
are defined by
\bes 
      M_i = \lim_{\mu\to\infty} (2b_0\gbar^2(\mu))^{-d_0/2b_0}\,\mbar(\mu)_i
\ees
and we have
\bes
     M_i/M_j = \mbar(\mu)_i/\mbar(\mu)_j = m^\mrm{bare}_i/m^\mrm{bare}_j\,.
\ees
%Clearly the RGI masses are the best choices for quark masses, but
%due to convention, one often converts to the $\msbar$-scheme, which is
%possible to high accuracy if $\mu$ is not too small. 

In \cite{mbar:pap1,mbar:nf2} the renormalization problem has been solved 
non-perturbatively in the $\nf=0,2$ theories: 
the $\mu$--dependence in the SF scheme was computed 
at the \% level. Based on this the results in \tab{t:masses}
were obtained. They may easily be converted to $\mbar_{\msbar}(\mu)$
if the desired renormalization scale 
$\mu$ is not too small.

\begin{table}[t!] \label{t:masses}
{%\footnotesize
\caption{Quark masses determined with full NP
        renormalization and continuum limit. We use
      $r_0=0.5\,\fm$.} 
\begin{center}
\begin{tabular}{lcccl}
\hline \\[-1.5ex]
  $i$     & $\nf$ &  input & $M_i/\GeV$ & ref. \\[0.5ex]
\hline\\[-1.5ex]
 strange & 0 & $\mk,r_0$ & 0.137(05) & \cite{mbar:pap3} \\
 strange & 2 & $\mk,r_0$ & 0.137(27) & \cite{mbar:nf2} \\
 charm & 0 & $\md,r_0$  & 1.654(45) & \cite{mbar:charm1} \\
 beauty & 0 & $\mbs,\mbsstar,r_0$ & 6.771(99) & \cite{lat05:nicolas} \\[0.5ex]     
\hline
\end{tabular}
\end{center}
\vspace{-8mm}
}
\end{table}

%%% Local Variables: 
%%% mode: latex
%%% TeX-master: "latticen.bib"
%%% End: 

For the charm and the beauty quark masses determinations with $\nf>0$
and NP renormalization are still missing. However, in our
opinion it is even quite early concerning the determinations of the 
light quark masses. Although some $\nf$-dependence of the strange 
quark mass has been reported in the literature, one can presently 
not exclude that this is due to perturbative uncertainties 
or discretization errors.

Note that the quark masses computed in the quenched approximation
were in a similar stage in 1996 but very soon 
afterwards the uncertainties shrunk by an order
of magnitude due to NP renormalization and continuum extrapolations! 

\bibliography{ll06}           %or whatever your .bib file is

\begin{thebibliography}{10}

\bibitem{alpha:sigma}
M. {L\"uscher}, P. Weisz and U. Wolff,
\newblock Nucl. Phys. B359 (1991) 221.

\bibitem{alpha:nf2}
ALPHA, M. Della~Morte et~al.,
\newblock Nucl. Phys. B713 (2005) 378, hep-lat/0411025.

\bibitem{alpha:letter}
ALPHA, A. Bode et~al.,
\newblock Phys. Lett. B515 (2001) 49, hep-lat/0105003.

\bibitem{pert:2loop_fin}
ALPHA, A. Bode, P. Weisz and U. Wolff,
\newblock Nucl. Phys. B576 (2000) 517, Erratum-ibid.B600:453,2001,
  Erratum-ibid.B608:481,2001, hep-lat/9911018.

\bibitem{mbar:pap1}
ALPHA, S. Capitani et~al.,
\newblock Nucl. Phys. B544 (1999) 669, hep-lat/9810063.

\bibitem{alpha:su2impr}
ALPHA, G. de~Divitiis et~al.,
\newblock Nucl. Phys. B437 (1995) 447, hep-lat/9411017.

\bibitem{alpha:su3}
M. {L\"uscher} et~al.,
\newblock Nucl. Phys. B413 (1994) 481, hep-lat/9309005.

\bibitem{4ferm:nf0}
ALPHA, M. Guagnelli et~al.,
\newblock JHEP 03 (2006) 088, hep-lat/0505002.

\bibitem{zastat:pap3}
ALPHA, J. Heitger, M. Kurth and R. Sommer,
\newblock Nucl. Phys. B669 (2003) 173, hep-lat/0302019.

\bibitem{lat01:jochen}
ALPHA, J. Heitger et~al.,
\newblock Nucl. Phys. Proc. Suppl. 106 (2002) 859, hep-lat/0110201.

\bibitem{pot:r0}
R. Sommer,
\newblock Nucl. Phys. B411 (1994) 839, hep-lat/9310022.

\bibitem{pot:intermed}
S. Necco and R. Sommer,
\newblock Nucl. Phys. B622 (2002) 328, hep-lat/0108008.

\bibitem{JLQCD:nf2b52}
JLQCD, S. Aoki et~al.,
\newblock Phys. Rev. D68 (2003) 054502, hep-lat/0212039.

\bibitem{QCDSF:nf2mstrange}
QCDSF, M. {G\"ockeler} et~al.,
\newblock (2004), hep-ph/0409312.

\bibitem{Bethke:2004uy}
S. Bethke,
\newblock (2004), hep-ex/0407021.

\bibitem{alpha:blum1}
J. {Bl\"umlein}, H. {B\"ottcher} and A. Guffanti,
\newblock Nucl. Phys. Proc. Suppl. 135 (2004) 152, hep-ph/0407089.

\bibitem{alpha:bernwetz}
W. Bernreuther and W. Wetzel,
\newblock Nucl. Phys. B197 (1982) 228.

\bibitem{alpha:Iwasaki1}
S. Takeda et~al.,
\newblock Phys. Rev. D70 (2004) 074510, hep-lat/0408010.

\bibitem{alpha:lepagemilk}
HPQCD, Q. Mason et~al.,
\newblock Phys. Rev. Lett. 95 (2005) 052002, hep-lat/0503005.

\bibitem{reviews:schlad}
R. Sommer,
\newblock (1997), hep-ph/9711243.

\bibitem{mbar:nf2}
ALPHA, M. Della~Morte et~al.,
\newblock Nucl. Phys. B729 (2005) 117, hep-lat/0507035.

\bibitem{mbar:pap3}
ALPHA, J. Garden et~al.,
\newblock Nucl. Phys. B571 (2000) 237, hep-lat/9906013.

\bibitem{mbar:charm1}
ALPHA, J. Rolf and S. Sint,
\newblock JHEP 12 (2002) 007, hep-ph/0209255.

\bibitem{lat05:nicolas}
M. Della~Morte et~al.,
\newblock PoS LAT2005 (2005) 224, hep-lat/0509173.

\end{thebibliography}
\bibliographystyle{h-elsevier2}   %if you use h-elsevier.bst

\end{document}